\documentclass[a4paper,prl,twocolumn,showpacs,preprintnumbers]{revtex4}

\usepackage{graphicx}
\usepackage{psfrag}
\usepackage{amsfonts,amssymb, amsmath}
\usepackage{dcolumn}
\usepackage{bm}

\newcommand{\ie}{i.e.\ }

\newcommand{\eps}{\varepsilon}

\newcommand{\ket}[1]{\,|#1 \rangle}

\newcommand{\ketbratext}[2]{\,|#1 \rangle\negthinspace\langle #2|\,}

\begin{document}

\title{Multipartite entanglement percolation}
\author{S. Perseguers$^1$, D. Cavalcanti$^2$, G. J. Lapeyre Jr.$^2$, M. Lewenstein$^{2,3}$, A. Ac\'in$^{2,3}$}
\affiliation{
    $^1$Max-Planck--Institut f\"ur Quantenoptik, Hans-Kopfermann-Strasse 1, 85748 Garching, Germany\\
    $^2$ICFO--Institut de Ci\`encies Fot\`oniques, Mediterranean Technology Park, 08860 Castelldefels, Spain\\
    $^3$ICREA-Instituci\'o Catalana de Recerca i Estudis Avan\c cats, Lluis Companys 23, 08010 Barcelona, Spain
}
\date{\today}

\begin{abstract}
We present percolation strategies based on multipartite measurements to propagate
entanglement in quantum networks. We consider networks spanned on
regular lattices whose bonds correspond to pure but non-maximally
entangled pairs of qubits, with any quantum operation allowed at
the nodes. Despite significant effort in the past, improvements over naive
(classical) percolation strategies have been found for only few lattices,
often with restrictions on the initial amount of entanglement in the bonds. In contrast,
multipartite entanglement percolation outperform the classical percolation protocols,
as well as all previously known quantum ones, over the entire
range of initial entanglement and for every lattice that we considered.
\end{abstract}

\pacs{03.67.Bg, 64.60.ah}
\maketitle


Quantum networks, which consist of neighboring nodes (or stations)
sharing entangled pairs of particles, are the future of quantum
communication~\cite{kimble}.
Establishment of entanglement between two nodes of the network
that are separated by an arbitrarily large distance represents the
most important task and challenge in the study of quantum
networks. The first solution to this problem was by means of
quantum repeaters \cite{briegel}, a (one-dimensional) line of
intermediate stations that share multiple noisy entangled
pairs. Long-distance entanglement is then achieved via entanglement
distillation at the repeater stations and the use of quantum memories.

A new approach for long-distance entanglement distribution was
introduced in~\cite{ACL07}, exploiting the connectivity of
multi-dimensional networks. In this setting networks are spanned on
regular lattices, where bonds are partially entangled pure states
of two qubits and with arbitrary quantum operations
allowed at the nodes. Using local operations and classical
communication (LOCC) one can probabilistically convert such states into perfect
singlets~\cite{vidal}. Therefore, the problem is solved by classical bond percolation:
entanglement propagation is possible when the conversion probability exceeds the
percolation threshold of the lattice~\cite{Stauffer,Grimmett}.
However, it has been shown in \cite{ACL07,PCA+08,LWL09} that this
\textit{classical entanglement percolation} (CEP) is
not optimal. In fact it is sometimes advantageous to consider
\textit{quantum entanglement percolation} (QEP): one precedes the
singlet conversion by a suitably designed set of local quantum operations,
and applies CEP afterwards. Despite several studies devoted to these strategies,
little is known about entanglement propagation protocols in general lattices.

In this Letter we introduce a powerful class of QEP protocols that
exploit multipartite entanglement. In contrast to the strategies
studied so far, which solely employ entanglement swapping and
conversion into singlets, we make full use of both the classical and quantum
aspects of quantum networks: connectivity and multipartite entanglement,
respectively. The interplay of geometrical lattice transformations
and entanglement manipulations is in fact a key ingredient to surpass CEP.
Before explaining our protocols in detail,
let us first state the main results of our work:
i) multipartite strategies relate for the first
time entanglement propagation to classical \emph{site
percolation} in a natural way~\cite{notesitebond};
ii) they systematically outperform all previously known classical
or quantum strategies, regardless of the initial entanglement of the bonds
and for every lattice that we considered;
iii) they are obviously applicable to any lattice relevant for physical problems.

Let us now be more specific and recall the main idea of CEP. Each bond of the
lattice is given by the state $\ket{\varphi} =
\sqrt{\varphi_0}\ket{00}+\sqrt{\varphi_1}\ket{11}$, with
$\varphi_0\geq\varphi_1$ and $\varphi_0+\varphi_1=1$, and can be optimally converted
into a singlet (or equivalently into the Bell pair
$\ket{\Phi^+}\propto\ket{00}+\ket{11}$), with
probability $p(\varphi)=2\varphi_1$. If
$p(\varphi)$ is larger than the threshold for the corresponding bond
percolation, then a giant cluster
of maximally entangled pairs appears and
long-distance entanglement between two nodes of this cluster is
achieved by performing entanglement swappings.

\begin{figure*}[!t]
   \psfrag{a}[][]{a)}\psfrag{b}[][]{b)}\psfrag{c}[][]{c)}
    \begin{center}
     \includegraphics[width=\linewidth]{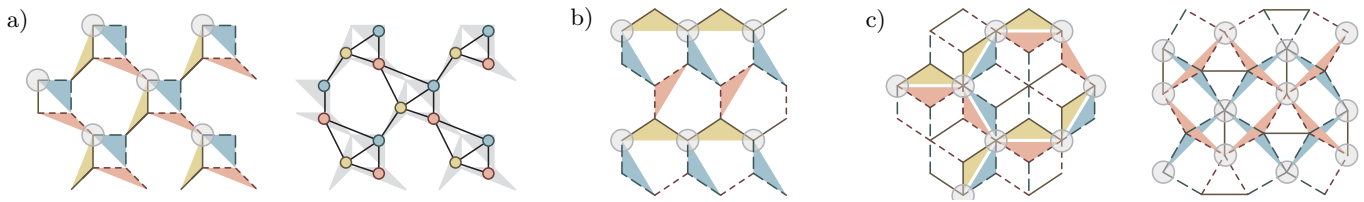}\\
     	\caption{(Color online)
     	From bond percolation to site percolation on a lattice with higher
		connectivity. First Step: Each measurement probabilistically
		transforms two links into a GHZ state on three
		nodes, which we depict by a triangle in $\mathcal{L}'$ and a
		site in $\hat{\mathcal{L}}$. Second step: two neighboring triangles
		can be merged into a bigger connected structure, which we represent
		by an edge in the transformed lattice.
            a) Measurement pattern to transform the $(4,8^2)$ lattice
            into a lattice with a better threshold. The right panel shows the
            corresponding site percolation problem;
            b) The hexagonal lattice is transformed into a
            lattice that can be denoted by
            $^1\!/\!_3(4,5^2)+\,^1\!/\!_3(3,4,5,4)+\,^1\!/\!_3(3,4,3,5^2)$;
            c) Measurement patterns
            for the dice and $(3^2,4,3,4)$ lattices, resulting in non-planar graphs (in the
            latter construction some links are converted into Bell pairs as in CEP).
            Semi-transparent circles highlight the nodes with highest coordination number
  			used for computing $\hat{\theta}$.}
    \label{fig:transforms}
    \end{center}
\end{figure*}

Concerning multipartite protocols, the main idea is to
produce Greenberger-Horne-Zeilinger (GHZ) states,
which are the generalization of the Bell pair $\ket{\Phi^+}$ to $n$ qubits:
$\ket{\text{GHZ}_n}\propto\ket{00\ldots0} + \ket{11\ldots1}$.
For instance, let us show how to build a tripartite GHZ state from
two neighboring bonds, that is, from two copies of $\ket{\varphi}$
sharing a common node. Guided by the optimality of Bell
measurements in the ZZ basis~\cite{notebasis} for one-repeater
entanglement swapping \cite[Sect.\,III\,A]{PCA+08}, we perform on
the two qubits of the common node a measurement $M$ consisting of the two
operators $\ketbratext{0}{01}+\ketbratext{1}{10}$ and
$\ketbratext{0}{00}+ \ketbratext{1}{11}$. 
The first outcome leads to a perfect GHZ state among the three nodes, but
with probability $\varphi_0^2+\varphi_1^2$ the second outcome
yields the (unnormalized) state
$\varphi_0\ket{000}+\varphi_1\ket{111}$. As for a pair of qubits,
the latter state is transformed into a GHZ state with
probability $2\varphi_1/(\varphi_0^2+\varphi_1^2)$. Summing the
two possibilities, we find that two copies of $\ket{\varphi}$ are
converted into a GHZ state of three qubits with optimal
probability $2\varphi_1$. The multipartite method derives its
power from the fact that, while the value of this probability is
the same as for the two-qubit swapping used in previous
entanglement percolation studies, we now have three entangled
qubits rather than two.

Although in this work we use only three-qubit GHZ states as
building blocks for multipartite entanglement percolation, we can generalize
the above procedure to construct GHZ
states of $n+1$ qubits starting from $n$ copies of $\ket{\varphi}$
sharing a common node. In this case we have
$2^{n-1}$ measurement operators $E_m$ of the form
$\ketbratext{0}{m} + \ketbratext{1}{\bar{m}}$, where $\bar{m}$ is
the complement of $m$ written in base 2. One can check that an
$(n+1)$-qubit GHZ state is then created with probability
\begin{equation}
    P(\text{GHZ}_{n+1}) = 1 - (\varphi_0-\varphi_1)\,\sum_{k=0}^{\left[\frac{n-1}{2}\right]}
        \binom{2k}{k}(\varphi_0\varphi_1)^{k},
    \label{eq:pGHZ}
\end{equation}
where $[x]$ denotes the integer part of $x$. This probability,
which is only optimal for $n\leq2$, can be understood as a sequence
of $n-1$ entanglement swappings in the ZZ basis on a chain of $n$
links, as calculated in \cite[Eq.~D3]{PCA+08}.

The multipartite strategy consists in creating a lattice $\hat{\mathcal{L}}$,
where nodes represent the GHZ states created from a lattice
$\mathcal{L}'$ \cite{notecircumflex}. Two vertices in $\hat{\mathcal{L}}$
are connected by a bond if the corresponding GHZ states share a 
common node in the original lattice, see Fig.~\ref{fig:transforms}a. This
defines a site percolation process with occupation probability $p=2\varphi_1$.
Above the percolation threshold of the new lattice, entanglement is propagated
over a large distance as follows. Consider the situation in which
two GHZ states of size $n$ and $m$ sharing one node have been
created. One builds a larger GHZ state on $n+m-1$ particles with
unit probability by applying the measurement $M$ on the two qubits
of the common node. This generalized entanglement swapping can be
iterated, and one eventually gets a giant GHZ state spanning
the network. Finally, note that, given a GHZ state of any
size, a perfect Bell pair is created between any two of its qubits
by measuring all other qubits in the $X$ basis.

At this point let us define the quantities characterizing long-distance entanglement.
Connectivity in percolation is characterized by the correlation length
$\xi(p)$ which approaches $0$ as $p\to0$ and as $p\to 1$, and diverges
as $p\downarrow p_c$ and as $p\uparrow p_c$.
In CEP we are interested in the probability $P'(A\leftrightarrow B)$
of creating a Bell pair between two nodes $A$ and $B$ separated by a distance $L$ and chosen
from the set of all nodes with the largest coordination number $Z_\text{max}$.
We shall assume that $A$ and $B$ satisfy $L\gg\xi'(p)$.
For $p<p'_c$, $P'(A\leftrightarrow B)$ decays exponentially in $L/\xi'(p)$, while for $p>p'_c$
the two nodes are connected only if they are both in $\mathcal{C}'$.
In this limit the events
$\{A\in\mathcal{C}'\}$ and $\{B\in\mathcal{C}'\}$ are independent,
which, together with translational invariance, gives
$P'(A\leftrightarrow B)=\theta'^2(p)$, where $\theta'(p)\equiv P'(A\in\mathcal{C}')$.
Thus the problem is reduced to studying $\theta'(p)$.
For QEP, we further restrict the two distant nodes to be chosen only from those
at which no measurement $M$ is to be made. In fact only these nodes retain the coordination number $Z_\text{max}$.
In terms of the probability measure on the site percolation process on $\hat{\mathcal{L}}$ we have
$\hat{\theta}(p)\equiv \hat{P}(\cup_i \{\hat{A}_i\in\hat{\mathcal{C}}\})$,
where the union is over all the $Z_\text{max}$ sites $\hat{A}_i$ possessing a qubit that is also in $A$.
It follows that $\hat{P}(A\leftrightarrow B)=\hat{\theta}^2(p)$,
and thus a direct comparison of $\theta'(p)$ and $\hat{\theta}(p)$ tells
us which from CEP or QEP is favorable. Note that this is the case
because in all our examples we have $p=p'=\hat{p}=2\varphi_1$.

Now that the basic ingredients have been presented, we
apply them to several natural examples of lattices. In the following paragraphs
we prove for eight lattices that multipartite entanglement
percolation protocols beats CEP. First, we examine the thresholds
and show that $\hat{p}_c<p'_c$ for each lattice. Then, we compute
the expansions $\hat{\theta}(p)$ and $\theta'(p)$ in the
high-density (\ie maximally-entangled) limit to prove that
$\hat{\theta}(p)>\theta'(p)$ as $p$ tends to unity. Finally, using Monte
Carlo techniques, we show that $\hat{P}(A\leftrightarrow
B)>P'(A\leftrightarrow B)$ for all $p\geq\hat{p}_c$.

\begin{table}
\begin{center}
    \begin{tabular}{c|@{\quad}c@{\qquad}c@{\qquad}c@{\qquad}c}
        \hline\hline\\[-.95em]
        \# & Lattice            & $p'_c$     & $\hat{p}_c$       & $\Delta$ [\%]\\
        \hline\\[-.95em]
        1   & $(4,8^2)$         & 0.6768    & 0.6499    & 4.0\\
        2   & Hexagonal         & 0.6527    & 0.609(0)  & 6.7\\
        3   & Kagom\'e          & 0.5244    & 0.427(1)  & 18.6\\
        4   & Square            & 0.5000    & 0.392(8)  & 21.4\\
        5   & Dice              & 0.4755    & 0.375(5)  & 21.0\\
        6   & $(3^2,4,3,4)$     & 0.4141    & 0.344(7)  & 16.8\\
        7   & Bowtie            & 0.4045    & 0.294(9)  & 27.1\\
        8   & Triangular        & 0.3472    & 0.273(5)  & 21.2\\
        \hline\hline
    \end{tabular}
    \caption{Improvement of entanglement percolation thresholds using a
    multipartite strategy, and relative gain $\Delta \equiv 1-\hat{p}_c/p'_c$.
    We performed Monte Carlo simulations to calculate $\hat{p}_c$
    for lattices 2--8; all other values can be found, with higher precision,
    in \cite{NMW08,Parviainen,Grimmett}.}
    \label{tab:pc}
\end{center}
\end{table}

We start by considering the Archimedean $(4,8^2)$ lattice,
see for instance \cite{SZ99} for the notation.
In Fig.~\ref{fig:transforms}a we propose a measurement pattern
indicating on which qubits the measurement $M$ is applied:
the new lattice
is denoted by $^2\!/\!_3(3^2,6^2)+\,^1\!/\!_3(3,6,3,6)$ and its
critical point is $\hat{p}_c\approx0.6499$ \cite{NMW08}. Since the
original threshold is $p'_c\approx0.6768$ \cite{SZ99}, the
proposed strategy yields an improvement over CEP.
It is interesting to note that the transformed lattice, arising from simple quantum
operations, is among the rather exotic examples considered in 
other studies. 
For example, this two-uniform lattice is considered in \cite{NMW08}, where a quantitative
relation between percolation thresholds and the Euler characteristic
is put in evidence.

We studied many other lattices and found that multipartite entanglement
percolation leads to better entanglement thresholds than
CEP in every case, which suggests that this may be a universal result.
To support this fact, let us describe seven
more well-known lattices and the transformations that improve
their critical points. In Fig.~\ref{fig:transforms}b--c we show the
measurement patterns for three of these lattices, namely the
hexagonal, dice and $(3^2,4,3,4)$ lattices. The four other
constructions are similar to the latter one: first, one or two
overlapping square lattices of double size are created from the
original lattice. Then, the remaining links are paired in a
regular way, as described in \cite[Fig.~14]{PCA+08} and
\cite[Fig.~7]{LWL09}. The key point of our constructions,
with respect to previous protocols,
is that we do not get disjoint lattices anymore, but rather
connected ones since the middle qubits can be used to
propagate entanglement through the network. We did not find
published critical values for site percolation
in the resulting lattices, mainly because they are non-planar or
non-regular graphs. Therefore, we turned to Monte Carlo
simulations to provide numerical results. The values obtained are summarized in
Tab.~\ref{tab:pc}: thresholds are not only better for all
lattices, but the gain is often significant, especially for
lattices of high connectivity.

\begin{table}
\begin{center}
    \begin{tabular}{c|@{\quad}c@{\qquad}c@{\qquad}c}
        \hline\hline\\[-.95em]
        \#  & $\theta'$                   & $\hat{\theta}$             & $f$\\
        \hline\\[-.95em]
        1   & $1-\eps^3-4\eps^4-11\eps^5$   & $1-\eps^3-4\eps^4-4\eps^5$    & $1/4$\\
        2   & $1-\eps^3-3\eps^4$            & $1-\eps^3-\eps^4$             & $1/4$\\
        3   & $1-\eps^4-6\eps^6$            & $1-\eps^4-2\eps^7$            & $1/3$\\
        4   & $1-\eps^4-4\eps^6$            & $1-\eps^4-4\eps^7$            & $1/2$\\
        5   & $1-\eps^6-6\eps^7$            & $1-\eps^6-9\eps^{10}$         & $3/4$\\
        6   & $1-\eps^5-5\eps^8$            & $1-\eps^5-\eps^8$             & $1/2$\\
        7   & $1-\eps^6-4\eps^8$            & $1-\eps^6-4\eps^{11}$         & $1/2$\\
        8   & $1-\eps^6-6\eps^{10}$         & $1-\eps^6-2\eps^{12}$         & $1/4$\\
        \hline\hline
    \end{tabular}
    \caption{Series expansions of $\theta'(p)$ and $\hat{\theta}(p)$, for $p=1-\eps$
    and $0\leq\eps\ll1$. These formulae have been derived for a fraction
    $f=\hat{d}/d'$ of nodes, where $d'$ ($\hat{d}$) is the density of nodes of higher
    connectivity in the original (transformed) lattice. Note that $d'=1$  for the
    Archimedean lattices, since by definition all their vertices are equivalent,
    while $d'=1/3$ for the dice and $d'=1/2$ for the bowtie lattice.}
    \label{tab:theta}
\end{center}
\end{table}

We provided numerical evidence that multipartite entanglement
percolation yields better thresholds than CEP, and therefore that
the connection probability between two widely separated nodes is increased
when the bond entanglement $p(\varphi)$ is small and lies in the
interval $(\hat{p}_c,p_c')$. We now
consider the opposite regime in which connections are
highly entangled, with $p=1-\eps$ and
$0\leq\eps\ll1$. In this situation an analytical study becomes
possible, and we use the perimeter method \cite[App.~C]{LWL09} to compute high-density
series expansions of $\theta'(p)$ and $\hat{\theta}(p)$. The
two lowest non-trivial orders of the expansions are easily calculated and are
sufficient to show that our strategy leads to an improvement over
CEP for all lattices, see Tab.~\ref{tab:theta}. Interestingly,
this improvement does not appear at the first non-trivial order
since there is no way to increase the number of independent
connections of a node. In fact generalized entanglement
swappings remove such connections, so that only the nodes where no quantum
operation is performed share the property $\hat{\theta}(p)>\theta'(p)$.

\begin{figure}
    \psfrag{t}[b][b]{$\hat{\theta}-\theta'$}
    \psfrag{p}[lb][]{$p$}
    \begin{center}
        \includegraphics[width=.9\linewidth]{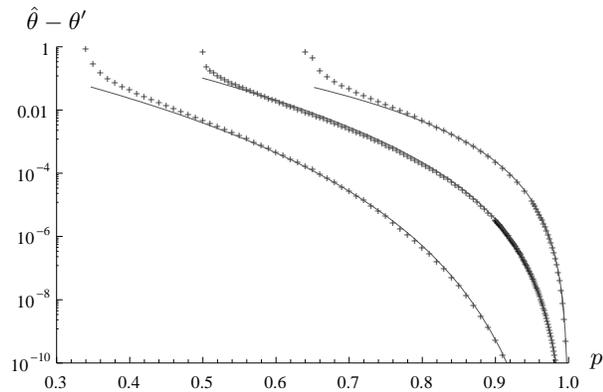}
        \caption{Numerical determination of  $\hat{\theta}-\theta'$ showing
        that multipartite percolation yields better long-distance
        entanglement for any initial entangled state $\ket{\varphi}$, with $p=2\varphi_1$.
        We show here the results for the triangle, square and hexagonal lattices
        (from left to right). High-density expansions are shown as light grey lines.}
    \label{fig:theta}
    \end{center}
\end{figure}

The previous analyses show that the proposed QEP strategy works
well near the classical thresholds and near the ideal situation
of perfect connections. We thus expect that our transformations
give $\hat{P}(A\leftrightarrow B)>P'(A\leftrightarrow B)$
for all values of bond entanglement. We performed Monte Carlo studies for the eight lattices
studied in this work, using importance sampling via the perimeter method in the high-density region,
attesting that $\hat{\theta}(p)$ is indeed strictly larger than $\theta'(p)$ for $p\in(\hat{p}_c,1)$.
This is shown for three lattices in Fig.~\ref{fig:theta}.

It is now legitimate to wonder about the optimality of our
multipartite entanglement schemes. It is not difficult to see that
they cannot be optimal for every value of the bond entanglement. In
fact, let us consider a square lattice and the deterministic ``centipede'' strategy
introduced in \cite[Sec. VI, A]{PCA+08}. The idea is that part of
the links are used to create a spiral centipede whose ``legs''
have length one. To that end, one performs two
entanglement swappings (measurements in the $XZ$
basis) on three links of a square. The resulting
state is combined with the fourth unused link to get a maximally
entangled state~\cite{vidal}. Clearly, this method can only be
applied to lattices whose nodes have four or more neighbors. For
square lattices, it leads to $f=1/2$ and $\theta(p)=1$ above the
point $\tilde{p}_c \approx 0.684$. Therefore, in the case of
highly entangled bonds, this methods is better than the present
protocol, which gives $\theta(p)=1$ only at the trivial point
$p=1$. However the centipede strategy fails whenever $p$
lies between $0.5$ and $\tilde{p}_c$, and therefore does not even
reach the classical threshold.

In conclusion, we have proposed a new approach to entanglement
percolation in quantum networks that makes significant improvements
over CEP. 
It makes use of multipartite entangled states to improve
the probability of long-distance entanglement generation
between nodes, regardless of the lattice topology or the degree
of entanglement of the bonds. In fact, we have provided both analytical and
numerical evidence for several well-studied lattices showing that
our protocols yield better results (lower thresholds and higher
probabilities) than all previously known strategies. At the
moment, we do not know if the thresholds obtained from our
multipartite QEP are optimal. This brings us to the important open
question of whether there is a minimum amount of initial
pure-state entanglement necessary for long-distance entanglement.
In one-dimensional chains we know that this is possible 
only if the connections are maximally entangled, but
while any percolation protocol
defines a sufficient condition for long-distance entanglement
distribution in higher-dimensional systems, a proof of the necessary amount of entanglement is
lacking.

Finally, the most important challenge is to generalize the results
obtained so far to the realistic case of mixed states. Recently,
some results have been presented in~\cite{BDJ09}, where bonds are
mixed states of rank two, which may be a first step in this
direction. However realistic noisy networks are best described by
full-rank mixed states, and even if we do not touch this problem
in this Letter, the generality of our results suggests that
multipartite strategies may also be of great importance in that case.


\begin{acknowledgments}
We acknowledge support from the QCCC program of the Elite Network
of Bavaria, the European QAP and PERCENT projects, the Spanish MEC
FIS2007-60182 and Consolider-Ingenio QOIT projects, Generalitat de
Catalunya, Caixa Manresa, and we thank Jan Wehr for helpful discussions.
\end{acknowledgments}

\bibliography{main}

\begin{thebibliography}{15}
\bibitem[{not({\natexlab{a}})}]{notesitebond}
\bibinfo{note}{Note that bond percolation can always be ``artificially'' mapped
  to site percolation.}
\bibitem[{not({\natexlab{b}})}]{notebasis}
\bibinfo{note}{We define a Bell measurement in the $ZZ$ or $XZ$ basis as a Bell
  measurement in which the local bases are the eigenvectors of the Pauli
  matrices $\sigma_Z$ and $\sigma_Z$, or $\sigma_X$ and $\sigma_Z$,
  respectively.}
\end{thebibliography}

\end{document}